\documentclass[fleqn,10pt]{wlscirep}
\usepackage[utf8]{inputenc}
\usepackage{multirow}
\usepackage[T1]{fontenc}
\title{BovineTalk: Machine Learning for Vocalization Analysis of Dairy Cattle under Negative Affective States }

\author[1]{Dinu Gavojdian}
\author[2]{Teddy Lazebnik}
\author[1]{Madalina Mincu}
\author[3]{Ariel Oren}
\author[1]{Ioana Nicolae}
\author[3,*]{Anna Zamansky}
\affil[1]{Cattle Production Systems Laboratory, Research and Development Institute for Bovine, Balotesti, Romania}

\affil[2]{Department of Cancer Biology, University College London, London, UK}

\affil[3]{Tech4Animals Lab, Information Systems Departrment, University of Haifa, Haifa, Israel}

\affil[*]{annazam@is.haifa.ac.il}

\begin{abstract}
There is a critical need to develop and validate non-invasive animal-based indicators of affective states in livestock species, in order to integrate them into on-farm assessment protocols, potentially via the use of precision livestock farming (PLF) tools. One such promising approach is the use of vocal indicators. The acoustic structure of vocalizations and their functions were extensively studied in important livestock species, such as pigs, horses, poultry and goats, yet cattle remain understudied in this context to date. Cows were shown to produce two types vocalizations: low-frequency calls (LF), produced with the mouth closed, or partially closed, for close distance contacts and open mouth emitted high-frequency calls (HF), produced for long distance communication, with the latter considered to be largely associated with negative affective states. Moreover, cattle vocalizations were shown to contain information on individuality across a wide range of contexts, both negative and positive. Nowadays, dairy cows are facing a series of negative challenges and stressors in a typical production cycle, making vocalizations during negative affective states of special interest for research. One contribution of this study is providing the largest to date pre-processed (clean from noises) dataset of lactating adult multiparous dairy cows during negative affective states induced by visual isolation challenges. Here we present two computational frameworks - deep learning based and explainable machine learning based, to classify high and low-frequency cattle calls, and individual cow voice recognition.  Our models in these two frameworks reached 87.2\% and 89.4\% accuracy for LF and HF classification, with 68.9\% and 72.5\% accuracy rates for the cow individual identification, respectively.
\end{abstract}
\begin{document}

\flushbottom
\maketitle
%
%
\thispagestyle{empty}


\section*{Introduction}

Farm animal welfare is commonly defined as the balance between positive and negative emotions, where positive emotions are considered as the main indicators of a moral animal life (‘a life worth living’ concept), with most of the recent research body of literature outlining the importance of affective states in farmed animals’ health and wellbeing \cite{webster2016animal,webb2019animal}. Non-human mammals’ affective states might vary in valence (positive to negative) and arousal levels (high to low), having functional adaptations linked to behavioural decisions that facilitate individual survival and reproduction, while promoting approaches towards rewards and avoidance \cite{kremer2020nuts,laurijs2021vocalisations}.

There is an evident need to develop valid non-invasive animal-based indicators of emotions in domestic animals in order to integrate them into future on-farm assessment protocols, potentially via the use of precision livestock farming (PLF) tools, such as novel sensors \cite{ben2016technology,matthews2016early}. To this end, the use of bioacoustics to evaluate health, emotional states, and stress responses has been validated for some of the most important livestock species such as pigs ({\em Sus scrofa domesticus}), goats ({\em Capra hircus}), horses ({\em Equus caballus}) and poultry ({\em Gallus gallus domesticus}). The research findings consistently show that vocal parameters differ substantially during positive and negative experiences \cite{silva2008cough,tallet2013encoding,whitaker2014sparse,briefer2015emotions,linhart2015expression,briefer2017perception,mcgrath2017hens,maigrot2018encoding,briefer2022classification}. Consequently, these developments started to be implemented and used in commercial settings in order to automatically classify animal vocalizations and identify health issues. For instance, the AI-based solution SoundTalks® was introduced in pig farms to detect respiratory diseases\footnote{https:// www.soundtalks.com/soundtalks/}. However, compared to the aforementioned species, there is a significant knowledge gap regarding cattle communication behaviour \cite{ede2019symposium,green2021vocal}. A potential explanation of this might be that cattle have a lower incidence of emitting vocalizations \cite{ede2019symposium,mcloughlin2019automated}, especially alarm and pain-specific vocalizations, developed as an adaptive response of the species as prey animals in order to avoid the risk of alarming potential predators. 

Domestic cattle vocalizations were shown to contain information on individuality, given the high levels of inter-cow variability in the acoustic characteristic of the vocalizations emitted under various contexts, as well as allowing facilitation of short- and long-distance interactions with herd-mates. This variability found in vocalizations produced by cattle allows for each animal to be identified by the ‘fingerprint’ of their call \cite{watts2001propensity,yajuvendra2013effective,de2015acoustic,green2019vocal}. Cattle are highly gregarious and form complex social relationships, having a strong innate motivation for continuous social contact \cite{boissy1997behavioral,holm2002calves}, with isolation from conspecifics resulting in physiological changes such as increased heart rates, cortisol levels, ocular and nasal temperature, and an increase in vocalization production \cite{muller2005behavioural,green2019vocal}. Furthermore, it was suggested that individual cattle vary in susceptibility to emotional stressors and challenges \cite{van2005responses,lecorps2018dairy,nogues2020individual}, with limited research being undertaken to evaluate the effects that isolation over prolonged periods of time has on vocalization response in adult cattle. All that said, we need to bear in mind that throughout a typical production cycle, dairy cows are facing a series of negative emotional challenges and stressors, such as separation from a calf immediately after calving, frequent regrouping based on production levels and lactation phase, re-establishing social hierarchy and dominance, frequent milking, isolation from herd-mates for insemination, pregnancy check-ups, being at high risks of developing metabolic disorders, isolation in sickness pens, to name a few. 

Cattle are known to produce two types of vocalizations, which are modulated by the configuration of the supra-laryngeal vocal tract \cite{de2015acoustic}. The first type are low-frequency calls (LF), produced by the animal with the mouth closed or partially closed, used for close distance contact, and regarded as indicative of lower distress or positive emotions. The second type are open-mouth emitted high-frequency calls (HF), produced for long-distance communication, and indicating higher arousal emotional states, generally associated with negative affective states \cite{briefer2012vocal,green2019vocal}. 

In domestic ungulates, individuality was proven to be encoded in a wide range of vocal parameters, most evidently in the F0-contour \cite{volodin2011nasal,briefer2022classification}, amplitude contour and call duration \cite{sebe2018amplitude}, as well as in filter-related vocal parameters including formant frequencies\cite{de2015acoustic}. Individuality expression was shown to be distinct for each call type \cite{volodin2011nasal,green2019vocal}, with individual differences in cattle high-frequency calls being attributed mainly to sound formants\cite{de2015acoustic,green2019vocal}, while vocalization formants are being modulated in turn by the caller vocal tract morphology \cite{taylor2016vocal}. Given that cattle can produce vocalizations with fundamental frequencies of over 1000Hz \cite{volodin2011nasal}, which are more likely to occur during times of higher arousal affective states, it was hypothesized by the authors that high-frequency calls encode a larger amount of individuality information, than their low-frequency equivalents, due to their propagation over longer distances where vision and/or olfactory signalling are not possible. 

Methods of studying animal vocal communication are becoming increasingly automated, with a growing body of research validating the use of both hardware and software that are capable of automatically collecting and processing bioacoustics data (reviewed by Mcloughlin et al.\cite{mcloughlin2019automated}). In this vein, Shorten and Hunter \cite{shorten2023acoustic} found significant variability in cattle vocalization parameters, and suggested that such traits can be monitored using animal-attached acoustic sensors in order to provide information on the welfare and emotional state of the animal. Therefore, automated vocalization monitoring could prove to be a useful tool in precision livestock farming \cite{mcloughlin2019automated,jung2021deep,li2021classifying}, especially as dairy farming systems become increasingly automated with wide-scale use of milking and feeding robots, all this having the potential to dynamically adjust the management practices while the number of animals per farm unit tends to increase.

Machine learning techniques are therefore increasingly applied in the study of cattle vocalizations. Some tasks that have been addressed to date include classification of high vs. low frequency calls \cite{shorten2023acoustic}, ingestive behaviour \cite{li2021classifying}, and categorization of calls such as oestrus and coughs \cite{jung2021deep}. 

This study makes the following contributions to the investigation of cattle vocalizations using machine learning techniques. First of all, we present the largest dataset to date of (n=20) cows’ vocalizations collected under a controlled ‘station’ setting, exclusively for negative affective states. Furthermore, we develop two types of AI models: deep-learning-based and explainable machine-learning-based for two tasks: (1) classification of high and low-frequency calls, and (2) individual cow identification. Finally, we investigate the feature importance of the explainable models.  

\section*{Methods}

{\em Ethical statement.} All experiments were performed in accordance with relevant guidelines and regulations. The experimental procedures and protocols were reviewed and approved by the Ethical Committee from the Research and Development Institute for Bovine, Balotesti, Romaina (approval no. 0027, issued on July 11, 2022), with the isolation challenge producing exclusively temporary distress to cows; the Ethical Committee of the University of Haifa waived
ethical approval.

\subsection*{Subjects and Experimental Approach}

The study was carried out at the experimental farm of the Research and Development Institute for Bovine in Balotesti, Romania. At the experimental facilities, cattle were managed indoors year-round (zero-grazing system), being housed under tie-stalls conditions (1.70/0.85 m, using wheat straws as bedding) in two identical animal barns, having access to outdoor paddocks (14-16 m$^2$/head) 10 hours/day, between milkings (7:00-17:00). Cows had ad libitum access to water and mineral blocks, receiving a daily ration of 30 kg corn silage, 6 kg of alfalfa hay and 6 kg of concentrates. In total, 20 lactating adult multiparous cows of Romanian Holstein breed were tested between August and September 2022. In order to avoid bias and to have a homogeneous study group, cattle included in our research were of similar age (lactations II \& III), were habituated previously to the housing system (min. 40 days in milk), and were comparable for body weight (619.5 $\pm$ 17.40 kg), which was important to avoid high variability for the vocal tract length, particularly on the larynx dimensions.  Cows were individually isolated inside the barns starting at 7:00 AM for 240 consecutive minutes post-milking, when the rest of the herd members were moved to the outside paddocks. After the commencement of recordings, animal caretakers were restricted from access to barns, and human traffic and machinery noise production outside the two barns were limited as much as possible. 

\subsection*{Vocalization recordings}
\begin{figure}[htb!]
\centering
\includegraphics[width=.8\columnwidth]{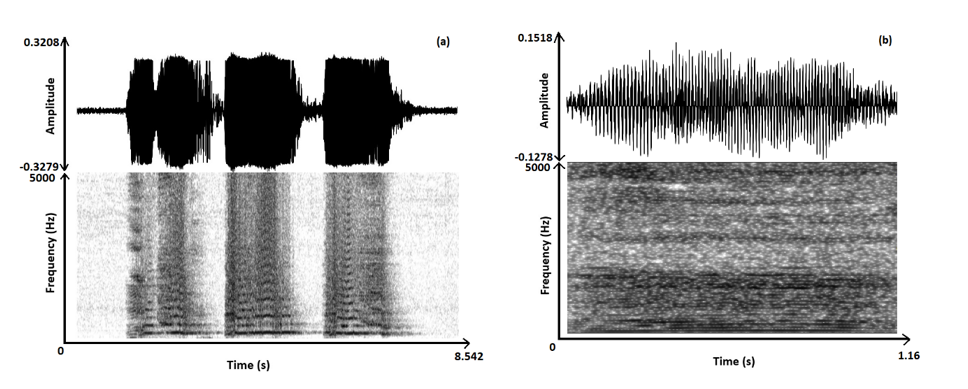}
\caption{Sequence of High Frequency Calls (HF) (a) and Low Frequency Calls (LF) (b) with oscillograms (above) and spectrograms (below) of typical vocalizations produced by cows during the isolation challenge.}
\label{fig:1}
\end{figure}

The vocalizations for this study were obtained using two identical directional microphones (Sennheiser MKH416-P48U3, frequency response 40-20.000 Hz, max. sound pressure level 130 dB at 1 kHz, producer Sennheiser Electronic®, Wedemark, Germany) attached to Marantz PMD661 MKIII digital solid-state recorders (with file encryption, WAV recording at 44.1/48/96kHz, 16/24-bit, recording bit rates 32-320 kbps, producer Marantz Professional®, UK). The microphones were directed towards the animal using tripods placed on the central feeding alleys at a distance of 5-6 m from the cows. For shock and noise reduction, Sennheiser MZW 415 ANT microphone windshields were used. After the end of each experimental day, vocal recordings were saved as separate files in the .WAV uncompressed format, at 44.1 kHz sampling rate and a 16-bit amplitude resolution. 
Despite the fact that all 20 multiparous lactating cows were isolated and recorded for 240 minutes post-milking, under identical conditions, not all cows vocalized with a similar frequency during the trials, resulting in the analysis of 1144 vocalizations (57.2 vocalizations per cow, ranging between 33 and 90 vocalizations per cow), out of which 952 were high-frequency vocalizations (HF) and 192 low-frequency vocalizations (LF). 
All sounds included in our investigation had undergone quality control check, while looking for clear, under- and un-saturated vocalizations, without combined environmental noises such as rattling equipment, chains clanging, or wind. Vocalizations were visualized on spectrograms using the fast Fourier transform method, at window lengths of 0.03 s, time steps of 1000, frequency steps of 250, dynamic range of 60 dB, and a view range between 0–5000 Hz (Figure \ref{fig:1}).

\begin{table}[htb!]
\centering
\begin{tabular}{p{0.2\textwidth}p{0.7\textwidth}}
\hline \hline
\textbf{Abbreviation / Unit of measure} & \textbf{Definition}                                                                                                                                    \\ \hline \hline
F0Mean (Hz)                             & Mean F0*   frequency value across the call                                                                                                             \\
F0Max (Hz)                              & Maximum F0   frequency value across the call                                                                                                           \\
F0Min (Hz)                              & Minimum F0   frequency value across the call                                                                                                           \\
F0Range (Hz)                            & Difference between minimum and maximum F0                                                                                                              \\
Q25 (Hz)                                & Frequency value at the upper limit of the first   quartiles of energy (below 25\%)                                                                     \\
Q50 (Hz)                                & Frequency value at the upper limit of the second quartiles   of energy (below 50\%)                                                                    \\
Q75 (Hz)                                & Frequency value at the upper limit of the third   quartiles of energy (below 75\%)                                                                     \\
Fpeak (Hz)                              & Frequency of peak amplitude                                                                                                                            \\
Sound Duration (s)                      & Duration of the call from start to end, measured on   the oscillogram                                                                                  \\
AMVar (dB/s)                            & Cumulative variation in amplitude divided by the   total call duration                                                                                 \\
AMRate (s-1)                            & Number of amplitude modulations in a certain time   frame                                                                                              \\
AMExtent (dB)                           & Mean-to-mean peak variation of each amplitude   modulation                                                                                             \\
Harmonicity (dB)                        & Degree of acoustic periodicity, also called   harmonic-to-noise ratio – higher values indicate more tonal voice                                        \\
F1Mean (Hz)                             & Mean frequency value of the first formant                                                                                                              \\
F2Mean (Hz)                             & Mean frequency value of the 2nd formant                                                                                                                \\
F3Mean (Hz)                             & Mean frequency value of the 3rd formant                                                                                                                \\
F4Mean (Hz)                             & Mean frequency value of the 4th formant                                                                                                                \\
F5Mean (Hz)                             & Mean frequency value of the 5th formant                                                                                                                \\
F6Mean (Hz)                             & Mean frequency value of the 6th formant                                                                                                                \\
F7Mean (Hz)                             & Mean frequency value of the 7th formant                                                                                                                \\
F8Mean (Hz)                             & Mean frequency value of the 8th formant                                                                                                                \\
Formant Dispersal (Hz)                  & Minimum spacing of the formants                                                                                                                        \\
Wiener Entropy Mean                     & Spectral flatness of a sound, calculated as the   ratio of a power  spectrum's geometric mean to its arithmetic mean   measured on a logarithmic scale \\ \hline \hline
\end{tabular}
\caption{Abbreviations and definitions of the 23 vocal parameters studied \cite{briefer2012vocal,de2015acoustic}.}
\label{table:1}
\end{table}

Vocalization recordings were then analyzed using Praat DSP package v.6.0.31 \cite{boersma2011praat}, as well as previously developed custom-built scripts \cite{reby2003anatomical,Beckers,briefer2015emotions,briefer2019expression,briefer2022classification}, for the automatic extraction of the 23 acoustic features of each vocalization, with the vocal parameters studied and their definitions being presented in Table 1, the output data being exported to Microsoft Excel for further analysis (see Supplementary material S.1).

\subsection*{Classification models}

We developed two different computational frameworks of the following types: 
\begin{enumerate}

\item Explainable model - a pipeline that uses as features the 23 vocal parameters described in Table 1, which have been studied in the context of cattle vocalizations. By using features which are highly relevant for our domain, we increase the explainability of our pipeline, allowing for the study of feature importance of our model; 
\item Deep learning model which uses learned features and operates as a “black-box”, that is not explainable. This model is expected to be more flexible and to have increased performance.  

\end{enumerate}
The \textit{explainable} framework is based on the TPOT \cite{le2020scaling}, AutoSklearn \cite{feurer2022auto} and H2O \cite{daugela2022real} automatic machine learning libraries. Namely, we assumed a dataset represented by a matrix \(x \in \mathbb{R^{n \times m}}\) and a vector \(y \in \mathbb{R}^n\), where \(n\) is the number of rows and \(m\) is the number of features in the dataset. Notably, we used the features described in Table 1, which makes the model more explainable, as the contribution of each feature to the model’s prediction could be computed. This dataset was divided into training and testing sets, such that the first have 80\% of the data and the latter the remaining 20\%, divided randomly. The training cohort was used to train the model and the testing cohort was used to evaluate its performance. Moreover, we randomly picked 90\% of the training dataset each time for r=50 times, making sure each value was picked at least half of the times. For each of these cohorts, we first obtain a machine learning pipeline from TPOT, AutoSklearn, and H2O aiming to optimize the following loss function: \(\sum_{i=1}^k \frac{a^i + f^i}{2k}\) where \(a^i\) and \(f^i\) are the \(i_{th}\) model’s instance accuracy and \(F_1\) scores, respectively, where k was the number of k-folds in the cross-validation analysis \cite{fushiki2011estimation}. Once all three models were obtained, we used all three of them to generate another cohort containing their predictions and the corresponding y value. These were then used to train an XGboost \cite{chen2016xgboost} model for the final prediction. For the hyperparameter tuning of the XGboost, we took advantage of the grid-search method. Finally, a majority vote between the \(r\)instances were used to determine the final model’s prediction. We reported the results for k=5 fold cross-validation over the entire dataset. 

The DL framework was adopted from Ye and Yang\cite{ye2021deep} which proposed a deep-gated recurrent unit (GRU) neural network (NN) model, combining a two-dimensional convolution NN and recurrent NN based on the GRU cell unit that gets as input the spectrogram of the audio signal. Generally, the two-dimensional convolution NN is used as a feature extraction component, finding spatio-temporal connections in the signal which than is being fed into the recurrent NN that operates as a temporal model able to detect short- and long- term connections in this feature space over time, these being effectively the rules for the voice identifications. For the hyperparameters of the model such as batch size, learning rate, optimization, etc., we adopted the values from Ye and Yang\cite{ye2021deep}. 

\newpage 
\section*{Results}

In this section, we examine the data obtained and outline the performance of the proposed explainable and DL models for the two different tasks based on the collected dataset. First, we provide a descriptive statistical analysis of the obtained dataset and its properties. Secondly, we present the performance of the models in classification between high and low-frequency calls. Finally, we present the models’ ability to identify each cow according to its vocalizations, divided into low, high, and all low + high vocalizations.
Table 2 summarizes the results of the explainable and DL models’ performance in separating between the high and low-frequency calls. The results are shown as mean ± standard deviation for k=5 fold. Importantly, we made sure that the train and test cohorts had vocalizations from both classes at each fold and that the ratio between the classes was kept between folds. Both models achieved good results with almost nine out of ten correct detections. One can notice that the DL model outperforms the explainable model. One explanation for this is that the DL is more expressive and therefore captures more complex dynamics which are not necessarily expressed by the features provided to the explainable model (see Table \ref{table:2}).

\begin{figure}[htb!]
\centering
\includegraphics[width=.8\columnwidth]{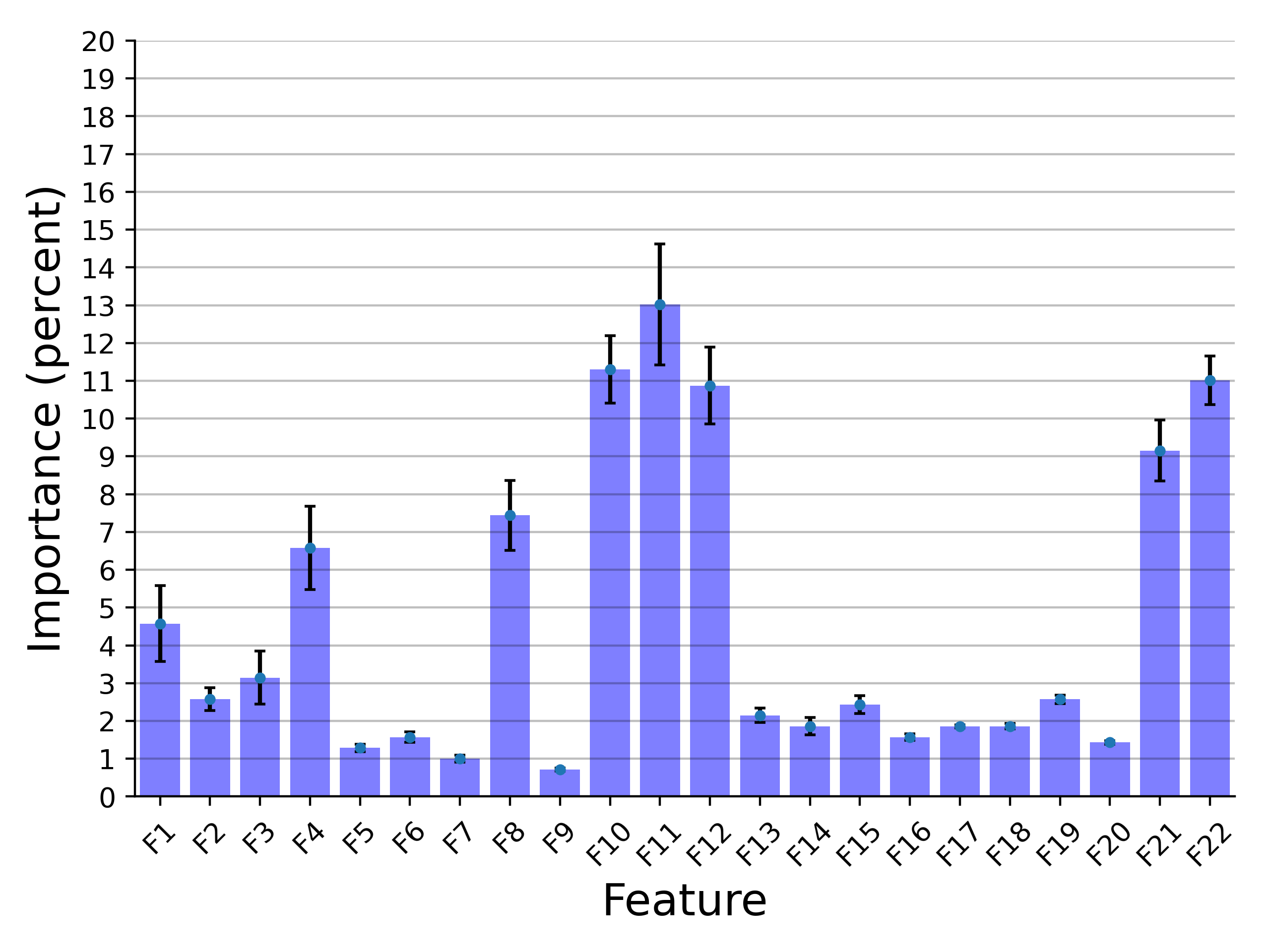}
\caption{The distribution of the features' importance for the \textit{high} and \textit{low} frequency calls (LF and HF) explainable classifier model. The results are shown as the average of a \(k=5\) fold cross-validation where the error bars indicate one standard deviation.}
\label{fig:2}
\end{figure}

Figure \ref{fig:2} presents the features’ importance of the explainable model for the high and low-frequency calls calculated by reducing one feature from the input and calculating its influence on the model’s performance. One can notice that AMvar, AMrate, AMExtent, Formant dispersal, and the Weiner entropy mean are the most important features, with joint importance of 55.36\%.

\begin{table}[htb!]
\centering
\begin{tabular}{lll} \hline \hline
Model       & Train set accuracy & Test set accuracy  \\ \hline \hline
Explainable & \(89.9 \pm 2.2\%\) & \(87.2 \pm 4.1\%\) \\ 
DL          & \(91.5 \pm 2.6\%\) & \(89.4 \pm 3.8\%\) \\ \hline \hline
\end{tabular}
\caption{The high and low frequency calls (LF and HF) classifier models’ performances. The results are shown as mean \(\pm\) standard deviation for \(k=5\) fold.}
\label{table:2}
\end{table}

\begin{figure}[htb!]
\centering
\includegraphics[width=.8\columnwidth]{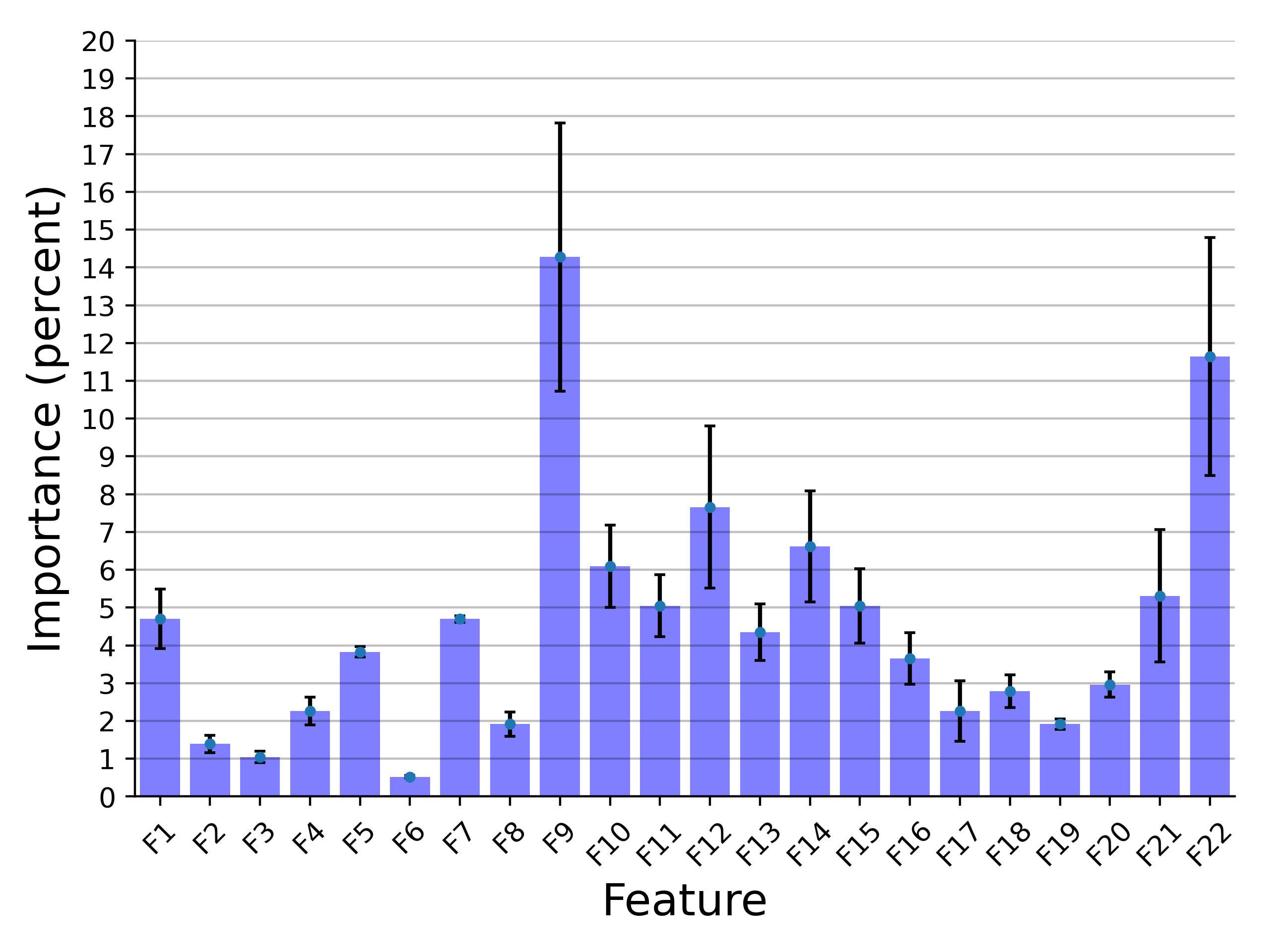}
\caption{The distribution of the features' importance for identification explainable classifier model for the Low Frequency + High Frequency dataset. The results are shown as the average of a \(k=5\) fold cross-validation, where the error bars indicate one standard deviation.}
\label{fig:3}
\end{figure}

\begin{table}[htb!]
\centering
\begin{tabular}{llll} \hline \hline
\textbf{Dataset}                               & \textbf{Model} & \textbf{Train set accuracy} & \textbf{Test set accuracy} \\ \hline \hline
\multirow{2}{*}{Low Frequency +High Frequency} & Explainable    &  \(73.0 \pm 3.3\%\)                            &     \(68.9 \pm 5.1\%\)                        \\
                                               & DL             &   \(76.3 \pm 4.2\%\)                           &         \(72.5 \pm 4.7\%\)                    \\
\multirow{2}{*}{Low Frequency}                 & Explainable    &          \(58.2 \pm 1.3\%\)                    &           \(50.9 \pm 2.8\%\)                  \\
                                               & DL             & \(65.5 \pm 1.8\%\)  &   \(46.8 \pm 3.3\%\)  \\
\multirow{2}{*}{High Frequency}                & Explainable    &   \(79.6 \pm 2.6\%\)                           &      \(68.4 \pm 3.2\%\)                       \\
                                               & DL             &  \(74.9 \pm 3.0\%\)                             &       \(70.8 \pm 3.4\%\)    \\ \hline \hline                 
\end{tabular}
\caption{The individual identification classifier models’ performances. The results are shown as mean \(\pm\) standard deviation for \(k=5\) fold.}
\label{table:3}
\end{table}

Table \ref{table:3} summarises the results of the explainable and DL models’ individual cow identification accuracy. The results are shown as mean ± standard deviation for k=5 fold. For this case, we made sure that the train and test cohorts had vocalizations from all cows such that the proportion of the vocalizations of each individual cow was present both in the train and test cohorts at each fold. The models obtained around 70\% accuracy, with a relatively low standard deviation. Like the previous experiment, the DL model outperforms the explainable model.

Figure \ref{fig:3} presents the features’ importance of the explainable model for the cow identification task, calculated by reducing one feature from the input and calculating its influence on the model’s performance. The sound duration played a critical role with 14.27\% importance, indicating that different cows have a significant pattern in their vocal duration, or at least a non-linear connection between the vocal duration to other features that allows for capturing unique identification patterns. The Wiener Entropy mean is the second best, with 11.65\% importance.

\section*{Discussion}
In this study, we present a dataset of cattle vocal recordings during negative affective states, which is, to the best of our knowledge, the largest dataset collected to date. The data from n=20 cows has been manually cleaned from background noises and trimmed to contain only the low-frequency (LF) and high-frequency (HF) calls, to ensure as higher quality of data as possible. The resulting dataset comprises 1144 records in total. Based on this data, we conducted two sets of tasks. Firstly, we provided a classifier for separating between low and high-frequency calls. Secondly, we provided a classifier for identifying individual cows based on their high-, low- or high + low-frequency vocalizations produced.

As shown in Table \ref{table:2}, both the explainable and DL models were able to accurately classify between the low- and high-frequency calls, with 87.2\% and 89.4\% accuracy, respectively. This outcome slightly outperforms (2\%, 4,4\%) the current state-of-the-art model \cite{shorten2023acoustic}, which used a smaller dataset of n=10 individuals. Notably, the differences between the models’ performances between the training and testing cohorts was around 2.5\% towards the training cohort, compared to the state-of-the-art which reports a 14.2\% difference. As such, our model is also less over-fitting, if at all. In addition, as the standard deviations of both models were 4.1\% and 3.8\%, this indicates that both models are robust.

For the individual cow identification task for both the LF and HF data, the explainable and DL models obtained 68.9\% and 72.5\% accuracy, respectively. When focusing only on the HF calls, the results were similar, with only 0.5\% and 1.7\% decrease in performance. On the other hand, when using only the LF samples, the accuracy sharply dropped to 50.9\% and 46.8\%, respectively, while also revealing overfit over the training dataset. This may be an indication that high-frequency calls contain more individuality information than low-frequency calls in cattle. These results are in accordance with previous findings across non-human mammals \cite{briefer2012vocal,briefer2019vocal}, where an increase in the arousal states was shown to lead to higher frequency vocalizations for both F0 and formant-related features, with vocalization parameters being more variable in negative- high arousal states.
An alternative explanation for this might be attributed to the reduced amount of LF data which contained 192 samples (i.e., 16.8\% of the entire dataset). While the performance of the model of Shorten and Hunter \cite{shorten2023acoustic} was better, this study worked with a reduced dataset for LF calls. In addition, their results may be indicative of overfitting, while explainable frameworks were not considered. 

Considering homologies in the physiology of vocalization production and the commonalities found across species \cite{briefer2019vocal}, the current findings could be extrapolated to other related species such as water-buffalo and Bos indicus species. To support this statement, Maigrot et al. \cite{maigrot2018encoding} found the functions of vocalizations to exceed intraspecies exchanges of information in domestic horses and Przewalski’s horses, wild boars and domestic pigs, these species being able to discriminate among positive and negative vocalizations produced by heterospecifics, including humans. Moreover, another potential contribution of the current research becomes apparent based on the experimental design and data collection. Whereas the studies conducted on cattle communication behaviour up-to-date analyzed predominantly vocalizations emitted by cows either in an un-controlled setting (e.g., mob on pasture or inside the barn), or assessed and compared calls among a wider set of contexts (e.g., positive and/or negative, with different putative valences and arousal levels), our setting was exclusively focused on a single negative context, with changes in affective states of the same animal being proven previously to result in modulations of the vocal parameters \cite{briefer2012vocal,lavan2019flexible,briefer2019expression}. 

Our results are in alignment with previous research which showed that isolation from herd-mates induces a wide range of behavioural and physiological responses in cattle \cite{muller2005behavioural,green2019vocal}, given the much higher incidence of HF calls observed during the isolation challenge, and taking into account the previous research results which suggested that the production and broadcasting of a repetitive single call type is indicative of persistent negative affective states \cite{collier2017call}, while reflecting a high urgency for the animal itself \cite{kershenbaum2016acoustic}.

This research is, however, not without its limitations. Factors such as emotional contagion among herd-mates, and thus the potential biological role of the distress vocalizations emitted by cows during the isolation challenge was not studied in the current trial. To address this, in our future research, we plan to include the use of additional sensors such as heart rate monitors, infrared thermography and stress-related biomarkers, to have a more generalized approach when evaluating emotional response to negative contexts. Moreover, considering the psychology and behavioural patterns of the species, mental processes such as learned helplessness could have contributed to the time-modulation of the vocal parameters following herd isolation, with animals abandoning their attempts to signal the negative event due to a perceived lack of control, which, however, does not mean that the negative event is being perceived as neutral by the animals. Additionally, the study herd consisted of multiparous adult cows, with various degrees of existing habituation to social isolation being presumed.

To summarize, cattle vocalizations can be seen as commentaries emitted by an individual on their own internal affective state, with the challenges lying in understanding and deciphering those commentaries. Looking forward, significantly more work needs to be done, taking into account a wider range of contexts and potential influencing factors on the vocal cues, in order to be able to draw strong conclusions regarding arousal or valence in cattle bioacoustics. Our study highlights the promising applications of machine learning approaches in cattle vocalization behaviour.

\section*{Conclusion}

In this study, we compiled a dataset consisting of cattle vocal recordings during negative affective states, which is one of the largest and cleanest datasets of its kind. Through our experiments using explainable and DL models, we have demonstrated the effectiveness of these models in classifying high- and low-frequency calls, as well as for identifying individual cows based on their vocalization productions. These results highlight the future potential of vocalization analysis as a valuable tool for assessing the emotional valences of cows and for providing new insights into promoting precision livestock farming practices. By monitoring cattle vocalizations, animal scientists could gain crucial insights into the emotional states of the animals, empowering them to make informed decisions to improve the overall farm animal welfare. Future work endeavours can take these results a step forward, gathering cattle vocalizations at different critical affective states to identify possible health risks or early real-time disease diagnostic.

\section*{Acknowledgements}

This work was supported by a grant of the Ministry of Research, Innovation and Digitization, CNCS - UEFISCDI, project number PN-III-P1-1.1-TE-2021-0027, within PNCDI III.

\section*{Author contributions statement}

DG designed methodology, obtained funding, provided resources and project administration; DG, IN \& AZ coordinated the work; MM collected and analysed the bioacoustics data; GD, TL, AO \& AZ analysed and interpreted the results; MM \& AO conceptualised the manuscript; GD, TL, IN \& AZ performed manuscript revision. All authors contributed critically to the writing of the manuscript and gave final approval for publication.

\section*{Data availability}

The datasets generated and/or analysed during the current study are available in the GitLab repository at this \href{https://gitlab.com/is-annazam/bovinetalk}{link}.

\bibliography{sample}

\end{document}